\documentclass{ws-p8-50x6-00}

\begin{document}

\title{Can the inflaton and the quintessence scalar be the same field?}

\author{Francesca Rosati}

\address{International School for Advanced Studies\footnote{Address from January
1st 2000:  Centre for Theoretical Physics, University of Sussex, Falmer,
Brighton BN1 9QJ, UK.  E-mail:  rosati@pcss.maps.susx.ac.uk.}  \\ via Beirut
2-4, 34014 Trieste, ITALY\\ E-mail:  rosati@he.sissa.it}


\maketitle

\abstracts{The most recent cosmological data point in the direction of a
cosmological constant dominated universe.  A suitable candidate for providing
the required acceleration is a rolling scalar field
named ``quintessence''.  We address the issue of building a particle physics
model for quintessence in the context of Susy QCD.  We then go on to ask if the
quintessence scalar can be identified with the inflaton field and give two
examples in which this is indeed the case.}

\section{Introduction} 
The very last years have witnessed growing interest in
cosmological models with $\Omega_m \sim 1/3$ and $\Omega_\Lambda \sim 2/3$,
following the most recent observational data (see for example the
discussion in the paper by Bahcall {\it et al.} \cite{data} and 
references therein).  
A very promising candidate for a dynamical cosmological constant 
is a rolling scalar field, named ``quintessence''. \cite{quint} 
The main motivation for constructing such dynamical schemes resides in the 
hope of weakening the fine tuning issue implied by the smallness of $\Lambda$.
In this respect, a very suitable class of models is provided by inverse power
scalar potentials which admit attractor solutions 
characterized by a negative equation of state. \cite{scalcosmo,track}

Consider the cosmological evolution of a scalar field $Q$, with potential
$V(Q)=M^{4+\alpha }Q^{-\alpha }$, $\alpha >0$, in a regime in which the scalar
energy density is subdominant with respect to the background.  
Then it can be shown \cite{scalcosmo,track} that the solution 
$Q\sim t^{1-n/m}$, with $n=3(w_{Q}+1)$ and
$m=3(w_{B}+1)$, is an attractor in phase space.  
We have defined $w_Q$ to be the equation of state of the scalar field $Q$, 
and $w_B$ that of the background ($=1/3$ for radiation and $=0$ for matter).
The equation of state of the scalar field on the  attractor is
found to be $w_{Q}= (\alpha \, w_B -2)/(\alpha+2)$, which is always
negative during matter domination.  
As a consequence, the ratio of the scalar to background energy density is 
not constant but scales as $\rho _{Q}/\rho_{B}\sim a^{m-n}$, thus growing 
during the cosmological evolution, since $n$ $<m$.  
The behaviour of these solutions is determined by the cosmological background 
and for this reason they have been named ``trackers'' in the 
literature. \cite{track}
A good feature of these models is that for a very wide range of the initial
conditions the scalar field will reach the tracking attractor before the present
epoch. Depending on the initial time, you can have several tens of 
orders of magnitude of allowed initial values for the scalar energy density.
This fact, together with the negative equation of state,
makes the trackers feasible candidates for explaining the cosmological
observation of a presently accelerating universe.
The point at which the scalar and matter energy densities are of the same order 
depends on the mass scale in the potential.  This is fixed by requiring that
$\Omega_Q = {\cal O} (1)$ today.

An interesting question, then, is whether the `quintessence' scalar and the
inflaton field, which dominate the expansion of the universe at very different
times, could indeed be the same field.  
If this is the case, it should also be possible to uniquely fix the 
initial conditions for the `quintessential' rolling from the end of inflation.
Models in which one single scalar field drives both inflation and the late 
time cosmological accelerated expansion are named ``quintessential inflation'' 
models. \cite{pv,noi2}

\section{A particle physics model: Supersymmetric QCD} 
As first noted by Bin\`{e}truy \cite{bin}, supersymmetric QCD theories with 
$N_{c}$ colors and $N_{f}<N_{c}$ flavors may give an explicit 
realization of a quintessence model with an inverse power law scalar potential.
The matter content of the theory is given by the chiral superfields $Q_{i}$ and
$\overline{Q}_{i}$ ($i=1\ldots N_{f}$) transforming according to the $ N_{c}$
and $\overline{N}_{c}$ representations of $SU(N_c)$, respectively.  In the
following, the same symbols will be used for the superfields $Q_{i}$,
$\overline{Q}_{i}$, and their scalar components.
Supersymmetry and anomaly-free global symmetries constrain the superpotential 
to the unique {\it exact} form 
\begin{equation} 
W=(N_{c}-N_{f})\left[ \Lambda ^{(3N_{c}-N_{f})} / \,
{\rm det}\, T \right] ^{ \frac{1}{N_{c}-N_{f}}} 
\label{superpot}
\end{equation} 
where the gauge-invariant matrix superfield $T_{ij}=Q_{i}\cdot
\overline{Q}_{j}$ appears.  $\Lambda $ is the only mass scale of the theory.

We consider \cite{noi} the general case in which different initial conditions are assigned
to the different scalar VEV's $\langle Q_{i}\rangle = \langle
\overline{Q}_{i}^{\dagger}\rangle \equiv q_i$, and the system is described by
$N_{f}$ coupled differential equations. 
In analogy with the one-scalar case, we look for power-law solutions of the form
\begin{equation}
q_{tr,i}=C_{i}\cdot t^{\, p_{i}}\ , \ \ i=1,\cdots ,\ N_{f}\ .
\label{scaling}
\end{equation}
It is straightforward \cite{noi} to verify that for fixed
$N_f$ (and when $\rho _{Q}\ll \rho _{B}$), a solution exists with $p_i \equiv p
= p(N_c)$ and $C_i \equiv C =C(N_c,\Lambda)$ and that it is the same for all the $N_f$
flavors.  
The equation of state of the tracker is given by
\begin{equation} 
w_{Q}=\frac{1+r}{2}w_{B}-\frac{1-r}{2}\ , \label{eosfree}
\end{equation} 
where we have defined $r \equiv N_{f}/N_c$.
Then, even if the $q_{i}$'s start with different initial
conditions, there is a region in field configuration space such that the system
evolves towards the equal fields solutions (\ref{scaling}), and the late-time
behavior is indistinguishable from the case considered by Binetr\'uy \cite{bin}
where equal initial conditions for the $N_f$ flavors were chosen.  
In spite of this, the multi-field dynamics introduces some new 
interesting features.  
For example, we have found that (in the two--field case) for any given 
initial energy density such that, for $q^{in}_1/q^{in}_2 =1$, the tracker 
is joined before today, there exists always a limiting value for the 
fields' difference above which the attractor is not reached in time.  
A more detailed discussion and numerical results about the
two-field dynamics can be found in Masiero {\it et al}. \cite{noi}

\section{Quintessential inflation} 
As already discussed, the range of initial conditions which allows 
$\rho_Q$ to join the tracker before the present epoch is very wide.  
Nevertheless, it should be noted that in principle we do not have any 
mechanism to prevent $\rho_Q^{in}$ from being outside the desired interval.  
In this respect, an early universe mechanism which could naturally 
set $\rho_Q^{in}$ in the allowed range for a late time tracking, would be 
highly welcome.
Moreover, if we require the quintessence scalar to be identified with the 
inflaton, we would at the same time obtain a tool for handling the 
initial conditions and a simple unified picture of the early and 
late time universe dynamics, which we call ``quintessential 
inflation''. \cite{pv,noi2}

The basic idea is to consider inflaton potentials which, as 
it is typical in quintessence, go to zero at infinity like inverse 
powers.  
In this way it is possible to obtain a late time quintessential behaviour 
from the same scalar that in the early universe drives inflation.  
The key point resides in finding a potential which satisfies the condition 
that inflation and late time tracking both occur, 
and that they occur at the right times (see Peloso 
{\it et al.} \cite{noi2}).

Two models have been shown to fulfill these two requirements.
One example \cite{noi2} is a first-order inflation model with potential going  
to zero at infinity like $\phi^{-\alpha}$. A bump  at  $\phi
\ll M_p$ allows for an early stage of inflation while the scalar  field
gets ``hung up'' in the metastable vacuum of the theory.  Nucleation of
bubbles of true vacuum through the potential barrier  sets the end of
the accelerated expansion and starts the reheating phase.
After the reheating process is completed, the quintessential  
rolling of the scalar $\phi$ starts and its initial conditions 
(uniquely  fixed by the end of inflation) can be shown to naturally 
be within the range which leads to a present day tracking.
As an alternative, we considered \cite{noi2} the model of hybrid 
inflation proposed by Kinney {\it et al}. \cite{riotto}
This is shown to naturally include a late--time quintessential behavior. 
As typical of hybrid  schemes, the potential at early times (that is 
until the inflaton field is smaller than a critical value $\phi_c\,$) 
is dominated by a constant term and inflation takes place. 
Eventually the inflaton rolls above $\phi_c\,$, rendering unstable the
second scalar of the model, $\chi$. 
This auxiliary field starts oscillating about its minimum and in this stage the 
universe is reheated. 
After $\chi$ has settled down, the inflaton continues its slow
roll down the residual potential, which goes to zero at infinity like
$\phi^{-2}$, thus allowing for a quintessential tracking solution. 
Also in this case the initial conditions for the quintessential 
part of the model are not set by hand, but depend uniquely on
the value of the inflaton field at the end of reheating.

\section*{Acknowledgments} I would like to thank Antonio Masiero, Massimo
Pietroni and Marco Peloso with whom I obtained the results presented in this
talk.

\end{document}